\documentclass[a4paper,prb,aps,twocolumn,showpacs,superscriptaddress]{revtex4-1}
\usepackage{amsmath}
\usepackage{amssymb}
\usepackage[greek,english]{babel}
\usepackage{graphicx}
\usepackage{bm}
\newcommand{\ket}[1]{\left|#1\right\rangle}

\newcommand{\braket}[3]{\left\langle #1\middle|#2\middle|#3\right\rangle}
\newcommand{\inprod}[2]{\left\langle #1\middle|#2\right\rangle}
\newcommand{\expect}[1]{\left\langle #1 \right\rangle}

\begin{document}
\latintext
\author{Erik Welander}
\affiliation{Department of Physics, University of Konstanz, Germany}

\author{Evgeny Chekhovich}
\affiliation{Department of Physics and Astronomy, University of Sheffield, United Kingdom}

\author{Alexander Tartakovskii}
\affiliation{Department of Physics and Astronomy, University of Sheffield, United Kingdom}

\author{Guido Burkard}
\affiliation{Department of Physics, University of Konstanz, Germany}

\title{Influence of Nuclear Quadrupole Moments on Electron Spin Coherence in Semiconductor Quantum Dots}

\begin{abstract}
We theoretically investigate the influence of the fluctuating Overhauser field on the spin of an electron confined to a quantum dot (QD). The fluctuations arise from nuclear angular momentum being exchanged between different nuclei via the nuclear magnetic dipole coupling. We focus on the role of the nuclear electric quadrupole moments (QPMs), which generally cause a reduction in internuclear spin transfer efficiency in the presence of electric field gradients. The effects on the electron spin coherence time are studied by modeling an electron spin echo experiment. We find that the QPMs cause an increase in the electron spin coherence time and that an inhomogeneous distribution of the quadrupolar shift, where different nuclei have different shifts in energy, causes an even larger increase in the electron coherence time than a homogeneous distribution. Furthermore, a partial polarization of the nuclear spin ensemble amplifies the effect of the inhomogeneous quadrupolar shifts, causing an additional increase in electron coherence time, and provides an alternative to the experimentally challenging suggestion of full dynamic nuclear spin polarization.
\end{abstract}

%
% 73.21.La 	Quantum dots 
%
% 71.70.Jp	Nuclear states and interactions
%
% 74.25.nj	Nuclear magnetic resonance
%
% 76.60.Lz	Spin echoes
%
% 76.60.Es	Relaxation effects
%
% 76.60.Gv	Quadrupole resonance
%
% 71.55.Eq	III-V semiconductors
%
% 31.15.aj	Relativistic corrections, spin-orbit effects, fine structure; hyperfine structure
%
% 75.40.Gb	Dynamic properties (dynamic susceptibility, spin waves, spin diffusion, dynamic scaling, etc.)
\maketitle
\section{Introduction}
Using the spin of an electron confined to a quantum dot (QD) has been proposed as one possible implementation of a qubit\cite{lossdivincenzo}. One of the hardest challenges of its practical realization is the fast decoherence of the electron spin caused by its interaction with the effective, time-varying magnetic field known as the Overhauser field \cite{erlingsson, merkulov, khaetskii, fischer, petta1, coish1, fischer, coishbaugh, bayer2}. Physically, the Overhauser field originates from the hyperfine interaction between the electron spin and nuclear spins of the QD. The exchange of spin between different nuclei via dipolar coupling combined with an imhomogeneous hyperfine coupling strength lead to a time-varying Overhauser field. The loss of electron spin coherence can be partially avoided by applying a {\greektext p}-pulse at time $t = T/2$ causing a reversal of the electron spin propagation and leading to an electron spin echo at time $t = T$\cite{abragam,slichter, wang}. However, if the Overhauser field varies in the interval $[0,T]$, the electron spin state cannot be fully restored. Techniques to prolong the electron coherence time by reducing the fluctuations of the nuclear spins have been theoretically suggested\cite{economou, ribeiro, yao, witzel1, witzel2, lee,khod,uhrig,taka,stepanenko} and experimentally tested \cite{dzhioevkorenev, petta1, petta2, chekh2, bluhm1, bluhm2, laird, warburton, bayer3, bayer4}. In this paper, we study the effects of nuclear quadrupolar shifts which impede the transfer of nuclear spin by causing certain transitions to be energetically forbidden.

An atomic nucleus having a non-uniform charge distribution may posses an electric quadrupole moment\cite{slichter,abragam,blatt} which couples to electric field gradients (EFGs) causing a shift in energy, known as the quadrupolar shift. The EFGs can be external, originate from neighboring atoms not participating in the nuclear spin transfer processes, or due to strain. We focus on the special case for which the EFGs have in-plane symmetry and where the symmetry axis coincides with the axis of an externally applied magnetic field, $\mathbf B = B_0\hat z$. This leads to a quadrupolar shift in energy proportional to $I_z^2 + c$, where $\mathbf I = (I_x, I_y, I_z)^T$ is the nuclear spin operator and $c$ is a constant.

\begin{figure}
\includegraphics{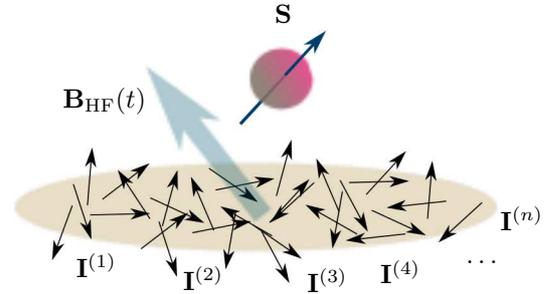}
\caption{\label{dotsketch}Illustration of the QD containing many nuclear spins (black arrows) each with corresponding operator $\mathbf I^{(n)}$. The nuclear spins couple to an electron spin (red ball with arrow) via the hyperfine Hamiltonian and give rise of an effective magnetic field (large blue arrow). Because of the transfer of nuclear spin between different nuclei and the inhomogenous hyperfine coupling strength, the effective magnetic field is fluctuating in time and is given by the stochastic vector $\mathbf B_{\rm HF}(t)$.}
\end{figure}

Recent experimental work\cite{chekh1} shows a significant increase in nuclear coherence times when quadrupolar energy shifts were introduced via strain. This suggests that the nuclear QPMs could be used as a way to prolong electron coherence times and provides an alternative to the experimentally challenging technique of complete dynamic nuclear spin polarization. In this paper, we try to estimate the effect of the nuclear QPMs on the electron spin coherence and its limitations.

In QDs, EFGs are primarily caused by strain\cite{guerrierharley, dzhioevkorenev, maletinsky, sinitsyn, slichter, bulutay, flis}, leading to displacements of the nuclei which in turn cause a modification of the charge distribution. If the nuclear displacement varies slowly over the QD and the QPMs are to a good approximation equal for all nuclei, the quadrupolar shift is homogenous and may be modeled by an additional term in the Hamiltonian which is equal for all nuclei. This may be the case when external strain is applied. If the stress is caused by a lattice mismatch at the interface between different materials (e.g. GaAs and InAs), the change in charge distribution will be more random, causing an inhomogenous quadrupolar shift that differs between different nuclear spins. In addition, the random location of dopants is another source of inhomogeneous quadrupolar shifts \cite{chekh1}. 
\begin{figure}[t]
\includegraphics[scale=1]{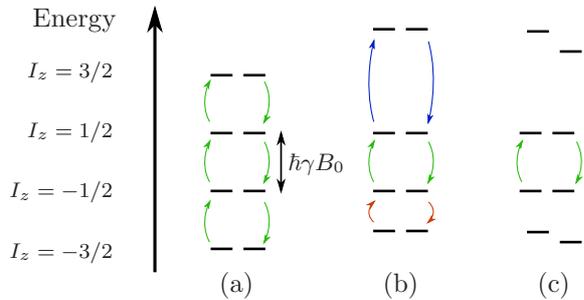}
\caption{\label{imgtrans}Energy level scheme of two nuclear spins $I = 3/2$ under the influence of an external magnetic field $B_0$ along $\hat z$ and quadrupolar shifts. The external magnetic field causes a Zeeman splitting of $\hbar\gamma B_0$ between spin levels differing by $\Delta I_z = 1$. (a) Without any quadrupolar shifts, any two neighboring energy levels differ by the same energy, thus allowing any transition for which the total nuclear spin along $\mathbf B$ is conserved. (b) With homogeneous quadrupolar shifts, all energy levels are shifted by an amount proportional to $I_z^2$ and transitions with different sets of initial and final states are inhibited. (c) With inhomogeneous quadrupolar shifts the energy levels of different nuclei are shifted by different amounts and only the $-1/2 \leftrightarrow 1/2$ transitions are energetically allowed. Here we have omitted any constant shift in energy, i.e. not depending on $I_z$, since it does not contribute to the nuclear spin dynamics.}
\end{figure}
\section{Theoretical model}
We study the dynamics of a single electron spin in a QD, containing $N$ atomic nuclei, each having spin $I$ in the presence of an external magnetic field, $\mathbf{B}$.
The electron and nuclear spins are influenced by each other via the hyperfine coupling, which we model with the Hamiltonian
\begin{equation}
\label{oheq}
H_{\rm{HF}} = \mathbf{S}\cdot\sum_{n = 1}^NA_n\mathbf{I}^{(n)},
\end{equation}
where $n$ enumerates the atomic sites, $A_n$ are hyperfine coupling strengths\cite{coishbaugh,coish1}, $\mathbf{S}$ is the electron spin operator and $\mathbf{I}^{(n)}$ are the nuclear spin operators. The hyperfine coupling strengths depend on the atomic species and are proportional to $|\Psi(\mathbf r_n)|^2$ where $\Psi(\mathbf r_n)$ is the electron envelope function at the atom site $n$ with the position $\mathbf r_n$. The nuclei are mutually coupled by their magnetic dipole moments\cite{slichter} as described by the Hamiltonian
\begin{equation}
\label{dieq}
H_{\textrm D} = \sum_{n < m}\alpha_{nm}\left(\frac{\mathbf{I}^{(n)}\cdot\mathbf{I}^{(m)}}{r_{nm}^3} - 3\frac{\left[\mathbf{I}^{(n)}\cdot\mathbf{r}_{nm}\right]\left[\mathbf{I}^{(m)}\cdot\mathbf{r}_{nm}\right]}{r_{nm}^5}\right),
\end{equation}
where $\alpha_{nm} = \gamma_n\gamma_m\hbar^2\mu_0/4\pi$ with the nuclear gyromagnetic ratios $\gamma_n$, $\mathbf{r}_{nm} = \mathbf{r}_n - \mathbf{r}_m$, and $r_{nm} = |\mathbf{r}_{nm}|$. In the presence of a strong magnetic field, the terms of Eq. (\ref{dieq}) not preserving the total nuclear spin projection along $\mathbf B$ are strongly suppressed. Assuming $\mathbf B = B_0 \hat z$ we can make the secular approximation
\begin{equation}
\label{seeq}
 H_{\textrm D^\prime} = \sum_{n < m }\alpha^\prime_{nm}\left[I^{(n)}_zI^{(m)}_z - \frac{1}{4}(I_+^{(n)}I_-^{(m)} + I_-^{(n)}I_+^{(m)})\right],
\end{equation}
where $\alpha^\prime_{nm} = \gamma_n\gamma_m\hbar^2(1-\cos^2\theta_{nm})/r_{nm}^3$ and $\theta_{nm}$ is the angle between $\mathbf{r}_{nm}$ and $\hat z$. The nuclear spins are further influenced by the electric quadrupole moments which, for the special case of planar symmetry and when the quadrupolar symmetry axis coincides with $\hat z$, can be modeled by the Hamiltonian
\begin{equation}
H_{\mathrm Q} = h\sum_n\nu_Q^{(n)} I_z^{(n)2},
\end{equation}
where we choose not to include any constant shift in energy since it would not affect the nuclear spin dynamics.
In principle this model could be used for several nuclear species at once, such as $^{69}$Ga, $^{71}$Ga, and $^{75}$As. However, different species typically have different gyromagnetic ratios and consequently have different spin transition energies. For this reason, the spin transfer between different nuclear species at high magnetic fields is strongly suppressed, and we include only one nuclear species.

The quantum state of the whole quantum dot including nuclear and electron spins is an element of the product Hilbert space $\mathcal H = \mathcal{H}_N\otimes\mathcal{H}_\textrm{e}$, where $\mathcal{H}_N = \mathcal{H}_I^{\otimes N}$ is the Hilbert space of the nuclear spins, $\mathcal{H}_I$ is the Hilbert space of one nuclear spin which is spanned by  $\{\ket{-I}, \ket{-I +1}\},\dots, \ket{I -1}, \ket{I}\}$, and $\mathcal{H}_\textrm{e}$ is the Hilbert space of the electron spin spanned by $\{\ket{\uparrow},\ket{\downarrow}\}$. In principle, the time evolution of any initial state $\ket{t = 0} \in \mathcal{H}$ is given by $\ket{t > 0} = e^{-i Ht/\hbar}\ket{t = 0}$ from the solution to the Schr\"odinger equation, where
\begin{equation}
H = H_{\rm{D}^\prime} + H_{\rm{Q}} + H_{\rm{HF}} + H_{\rm{Z}}
\end{equation} and
\begin{equation}
H_{\rm{Z}} = \hbar\mathbf{B} \cdot \left(\gamma_e \mathbf{S} + \gamma\sum_n\mathbf{I}^{(n)}\right)
\end{equation}
is the combined electron and nuclear Zeeman term with the electron gyromagnetic ratio $\gamma_e = g_e\mu_{\rm B}/\hbar$, where $g_e$ is the electron $g$-factor and $\mu_{\rm B}$ is the Bohr magneton. However, the dimension of the Hilbert space, $\dim \mathcal{H} = 2(2I + 1)^N$, grows exponentially with the number of nuclear spins $N$, and for a typical quantum dot containing $10^4$ to $10^6$ nuclei, a direct numerical calculation of its time evolution is unrealistic. To make a suitable approximation, we divide the problem into two parts by decoupling the electron from the nuclear spins. This allows us to consider a sample of fewer nuclear spins for which the spin dynamics are first simulated and then used as an input to the electronic problem.

To study the nuclear spin dynamics we consider a set of $M \ll N$ nuclei. From the eigenstates of $I_z^{(n)}$ for each nuclear spin we construct initial product states $\ket{\mathbf{m}_0,t = 0} = \ket{\mathbf{m}_0} \in \mathcal{H}_M$, where $\ket{\mathbf m} = \ket{m_z^{(1)}, m_z^{(2)}\dots m_z^{(M)}} = \ket{m_z^{(1)}}\otimes\ket{m_z^{(2)}}\otimes\dots\otimes\ket{m_z^{(M)}}$ and $m_z^{(n)}$ is the projection of the $n$-th nuclear spin along $\hat z$.
The product states are eigenstates of the total nuclear spin projection operator along $\hat z$, $I_z = \sum_{n = 1}^M I_z^{(n)}$ with eigenvalues $\sum_{n = 1}^M m_z^{(n)}$, and are evolved directly by $\ket{\mathbf m_0, t > 0} = e^{-i H^\prime t/\hbar}\ket{\mathbf m_0}$, where $H^\prime = H_{\rm D^\prime} + H_{\rm Q}$ is the Hamiltonian of the nuclear spins.
The time-evolved state vector gives the probability function for the eigenstates $\ket{\mathbf m}$ of $I_z$ as $p_{\mathbf m_0}(\mathbf m,t) = \left|\inprod{\mathbf{m}}{\mathbf{m}_0,t}\right|^2$ and from this probability function we define a stochastic vector $\mathbf m(t) = (m_z^{(1)}, m_z^{(2)} \dots m_z^{(n)})$ with probability $p_{\mathbf m_0}(\mathbf m,t)$. The effective magnetic field from any product state $\ket{\mathbf m}$ is given by
\begin{equation}
\mathbf{B}_{\rm{HF}}(\mathbf m) = B(\mathbf m)\hat z
\end{equation}
where
\begin{equation}
B(\mathbf m) = \braket{\mathbf m}{\sum_{n=1}^M A_n\mathbf{I}^{(n)}}{\mathbf m} = \sum_{n=1}^MA_nm_z^{(n)}.
\end{equation}
The hyperfine field $\mathbf B_{\rm HF}$ has vanishing components along $\hat x$ or $\hat y$ since $\braket{\mathbf m}{I_x^{(n)}}{\mathbf m} =   \braket{\mathbf m}{I_y^{(n)}}{\mathbf m} = 0$ for any $\mathbf m$ and $n$. Using the previously defined stochastic $\mathbf m(t)$, we obtain a discrete-valued stochastic magnetic field
\begin{equation}
B(t) = B(\mathbf m(t))
\end{equation} 
with non-Markovian dynamics. Although the probability function of $p_{\mathbf m_0}(\mathbf m, t)$ is given by the time evolution for any $\mathbf m_0$, $B(t)$ is still a stochastic variable.
% and we may define a probability functional $p[B(t)] = \int_0^tp(B(t^\prime),t^\prime)\,dt^\prime$ for each possible realization $B(t)$ of $B(t)$.
%, where for any time $t$, $\mathbf{b}(t) = \mathbf{b}(\mathbf{m}(t)) \in\left\{\sum_{n=1}^MA_nm_z^{(n)}\hat{z} : m_z^{(n)} \in \{-I/2, -I/2 + 1, \dots I/2\}\right\}$.

For a given $B(t)$, finding the electron spin dynamics is straight-forward by considering the Hamiltonian
\begin{equation}
\label{eleq}
H_{\rm e}(t) = S_z (B(t) + \hbar\gamma_e B_0),
\end{equation}
which describes the time-evolution of an initial state by $\ket{t > 0} = e^{-i\int_0^tH_{\rm e}(t^\prime)\,dt^\prime/\hbar}\ket{t = 0}$.  The effects of the static magnetic field $B_0$ is completely cancelled by the electron spin echo and hence this term may be excluded from the dynamics. Formally this can be achieved by going over to the rotating frame\cite{slichter,abragam}.

In order to study the electron spin echo we let the electron spin state be given by $\ket{t} = c_\uparrow(t)\ket{\uparrow} + c_\downarrow(t)\ket{\downarrow}$ and choose $c_\uparrow(0) = c_\downarrow(0) = 1/\sqrt 2$. $H_{\rm e}(t)$ is diagonal in the eigenbasis of $S_z$ and the time evolution is directly given by
\begin{equation}
\label{elevol}
\ket{t} = \frac{e^{-i\varphi(t)}\ket{\uparrow} + e^{i\varphi(t)}\ket{\downarrow}}{\sqrt 2},
\end{equation}
where $\varphi(t) = \int_0^tB(t^\prime)\,dt^\prime/2$ and since we are using the rotating frame there is no extra phase difference from the electron Zeeman splitting. The change in electron spin state due to the evolving $\varphi(t)$ can be partially undone by applying a \greektext p\latintext-pulse around $\hat x$ at $t = T/2$ which transforms the electron state according to $\alpha\ket{\uparrow} + \beta\ket{\downarrow} \longrightarrow \beta\ket{\uparrow} + \alpha\ket{\downarrow}$ and at time $t = T$ the electron spin will be in the state $\ket{T} = (e^{2i\varphi(T/2)-i\varphi(T)}\ket{\uparrow} + e^{-2i\varphi(T/2)+i\varphi(T)}\ket{\downarrow})/\sqrt 2$. We denote the projection onto the initial electron state by
\begin{equation}
\lambda[B(t)](T) = \inprod{T}{0} = \cos[2\varphi(T/2) - \varphi(T)],
\end{equation}
which gives a measure of the quality of the electron echo as a function of echo time. An exhaustive description of the electron dynamics from a given initial nuclear state $\ket{\mathbf m_0}$ is given by averaging over all possible temporal realizations
\begin{equation}
f_{\mathbf m_0}(T) = \int\lambda[B(t)](T)p_{\mathbf m_0}[B(t)][DB(t)],
\end{equation}
where $p_{\mathbf m_0}[B(t)]$ is the probability density functional taking a function $B(t)$ as a parameter\cite{mandel}, and the $\int\dots[DB(t)]$ denotes the functional integration over all possible $B(t)$.
However, except for the case of the fully polarized initial state when $\mathbf m_0 = \pm (I, I, \dots, I)$, the set of all possible temporal realizations is infinite and since the $p(B(t),t)$ needs to be calculated numerically, $f_{\mathbf m_0}(T)$ is approximated by performing a set of random walks instead. For this purpose we let $B(t) = \sum_{n=1}^MA_nm_z^{(n)}(t)$ be given for a discrete set of times by randomly selecting $\mathbf m(t)$ with probabilities $p_{\mathbf m_0}(\mathbf m, t)$. This way we obtain the approximation
\begin{equation}
\tilde f_{\mathbf m_0}(T) = \frac{1}{K}\sum_{k=1}^K\lambda[B_k(t)](T),
\end{equation}
and $K$ is the number of samples and $B_k(t)$ are the randomly chosen realizations of the Overhauser field. This differs from a typical random walk of Monte-Carlo type since the steps are chosen from a time dependent probability distribution leading to non-Markovian dynamics.

A typical electron spin echo experiment consists of averaging several measurements for which the initial nuclear states do not need to be identical. To incorporate this we define an average fidelity for a set of $L$ measurements according to
\begin{equation}
\label{fieq}
F(T) = \frac{1}{L}\left(\sum_{l = 1}^L\tilde{f}_{\mathbf m_l}(T)\right)^2,
\end{equation}
where $\mathbf m_l$ represent the initial nuclear states, from which the probability distribution $p_{\mathbf m_l}(\mathbf m, t)$ is calculated numerically. The initial states are in turn chosen randomly with the thermal equilibrium probabilities
\begin{equation}
p(\mathbf m) = \frac{1}{Z}\exp\left[-\frac{\hbar\gamma B_0\sum_nm_z^{(n)}}{k_BT_N}\right],
\end{equation}
where we have used the partition sum
\begin{equation}
Z = \sum_{\mathbf m}\exp\left[-\frac{\hbar\gamma B_0\sum_nm_z^{(n)}}{k_BT_N}\right],
\end{equation}
and where $T_N$ is the nuclear spin temperature. We define the nuclear spin polarization as $\eta = -\expect{I_z}/NI$ with $\expect{I_z} = \sum_\mathbf{m}p(\mathbf m)\sum_{n = 1}^Nm_z^{(n)}$ which leads to the relation
\begin{equation}
\label{polareq}
\eta = \tanh\frac{g_N\mu_0B_0}{2k_BT_N}
\end{equation}
between $\eta$ and $B_0/T_N$.
\section{Results}
\begin{figure}[t]
\includegraphics[scale=1]{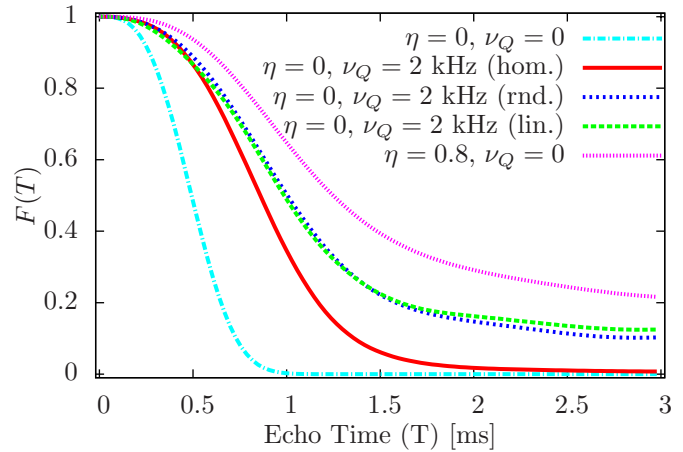}
\caption{\label{echoes1}Typical electron spin echo curves for various nuclear parameters. The cyan line shows the electron spin echo without quadrupolar moments and nuclear polarization. The red line shows the effect of including a homogeneous quadrupolar shift of $2$ kHz. The blue and green lines show the the effect of $2$ kHz random and linear inhomogeneous quadrupolar shift, where the inhomogeneity is characterized by a random distribution and linear gradient respectively, described below. The purple line shows the effect of nuclear spin polarization of $80$\%. We observe that for a highly polarized nuclear spin ensemble, as for the situation with large inhomogeneous quadrupolar shifts, the echo fidelity does not vanish completely even at long times.}
\end{figure}

Using ensembles of $6$ spins $I = 3/2$, arranged on a line with $\mathbf r_n = a n \hat x$ for $n = 1 \dots 6$ and $a = 5.56$ \AA, for each parameter set of polarization and quadrupolar shifts we performed $K = 10000$ random walks for each of $L = 1000$ random initial states producing typical fidelity vs. echo time curves shown in Fig. \ref{echoes1}. We used the hyperfine couplings $A_n = A e^{-n^2 / 6 ^2}$ to model the varying coupling strength for an electron in a QD. $A$ was adjusted to give a typical\cite{varwig} electron spin coherence time of $1$ ms for the unpolarized case and without quadrupolar shifts. Rather than in the absolute coherence time, we are primarily interested in the change of the coherence time due to polarization and quadrupolar shifts. 
%
%Although experiments\cite{chekh1,flis,bulutay,maletinsky} report shifts up to several MHz, we could not observe any significant change in electron spin coherence when exceeding $2$ kHz by several of orders of magnitude and thus we limit the quadrupolar shifts to $2$ kHz in our calculations.
Experiments\cite{chekh1,flis,bulutay,maletinsky,bayer1, sinitsyn} report QP shifts up to several MHz, and in initial calculations we investigated QP shifts in the MHz range. However, we found that the electron spin coherence does not change significantly when exceeding $2$ kHz, and thus we limit the quadrupolar shifts to $2$ kHz in our calculations.
In order to systematically study the effects of polarization and quadrupole moments, we fit the echo curve to the function
\begin{equation}
f(T) = (1 - F_\infty)\exp(-T^4 / T_2^4) + F_\infty,
\end{equation}
where $T_2$ will be called the coherence time and $F_\infty$ is an asymptotic value. Physically, the two terms can be regarded as the nuclear spin ensemble having both a fluctuating part causing the decaying term and a static one giving rise to the asymptote $F_\infty$. The form of the exponential $T^4$ decay can be found by considering low-frequency noise\cite{ithier}.
%For increasing polarization we would expect the static part to increase. 
%
\subsection{Effect of polarization}
\begin{figure}[t]
\includegraphics[scale=1]{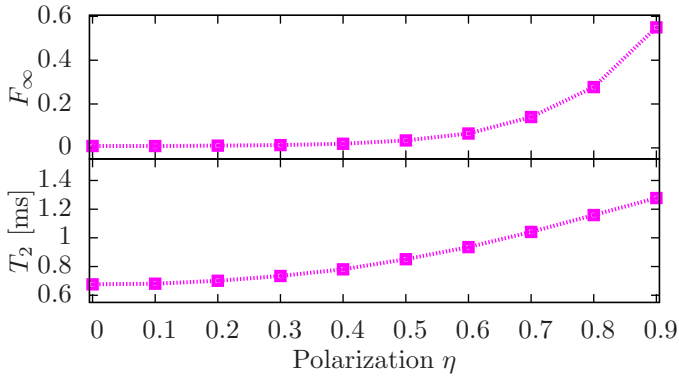}
\caption{Effect of increased nuclear spin polarization. When the nuclear spin polarization is increased, both electron coherence time $T_2$ and asymptotic fidelity $F_\infty$ increase. Not shown in the figure is the situation of $\eta = 1$, which would lead to a unit asymptotic fidelity or infinite electron coherence time.\label{pols1}}
\end{figure}

We begin with studying the effects of increasing the nuclear polarization without including quadrupole moments. Fig. \ref{pols1} shows the electron spin coherence time $T_2$ and the asymptotic fidelity $F_\infty$ as a function of nuclear spin polarization $\eta$. For increasing nuclear spin polarization both electron spin coherence time and fidelity asymptote increase. The increasing $F_\infty$ suggests that the nuclear spin dynamics is not only slowed down but also that there is a growing part of the nuclear spin ensemble that remains static. For a complete polarization $\eta = 1$ the nuclear spins become completely static and  $T_2 \rightarrow \infty$ and/or $F_\infty \rightarrow 1$. This is an expected result and polarizing the nuclear spins has been proposed as a method to prolong electron coherence times. Practically, this method has proven to be challenging and  so far $\eta = 65$\% is the the maximal dynamic nuclear spin polarization reported \cite{chekh2}, which further motivates searching for alternative ways to reduce the nuclear spin fluctuations.
\subsection{Effect of quadrupolar shifts}
\label{QPmoments}
We now turn our attention to the quadrupole moments. As described in the introduction, there is a significant difference between homogeneous quadrupolar shifts, where all nuclear spins experience the same shifts in energy, and inhomogeneous quadrupolar shifts, where each nuclear spin may experience a different effect. Using an unpolarized ensemble of $M = 6$ nuclear spins as before, the homogeneous quadrupolar shifts are modeled by
\begin{equation}
H_Q = h\nu_Q\sum_{n = 1}^M{I_z^{(n)}}^2.
\end{equation}
For the inhomogeneous quadrupolar shift we investigate two different distributions and use the Hamiltonian
\begin{subequations}
\begin{align}
\label{inhham}
H_Q =& \frac{2h\nu_Q}{M-1}\sum_{n = 1}^M(n-1){I_z^{(n)}}^2\\
\label{rndham}
H_Q =& \frac{h\nu_Q}{Y}\sum_{n = 1}^MX_n{I_z^{(n)}}^2,
\end{align}
\end{subequations}
where $X_n \sim $ U$(0,1)$ and $Y = \sum_{n=1}^M X_n / M$ , so that the average quadrupolar shift is $\nu_Q$ all cases. The Hamiltonian (\ref{inhham}) describes a linear gradient in quadrupolar shifts and Eq. (\ref{rndham}) describes random quadrupolar shifts.
\begin{figure}[b]
\includegraphics[scale=1]{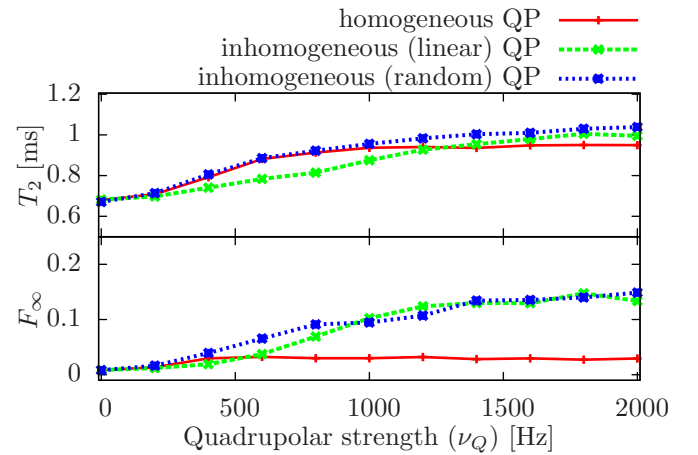}
\caption{\label{hom_vs_inh}Electron coherence time $T_2$ and asymptotic fidelity $F_\infty$ including homogeneous (red curves), linear (green curves, Eq. (\ref{rndham})), and random (blue curves, Eq. (\ref{rndham})) inhomogeneous QP shifts. \textit{Upper panel:} Electron coherence times $T_2$. For high QP strengths, the inhomogeneous shift leads to marginally longer electron coherence times since  all transitions except between $I_z = -1/2$ and $I_z = 1/2$ are energetically forbidden. For small QP strengths, the inhomogeneous shift may lead to a smaller change in electronic coherence times than for the homogeneous shift because parts of the nuclear system experience a relatively small shift in transition energy. \textit{Lower panel: }The asymptotic fidelity $F_\infty$ of the electron spin. At higher quadrupolar shifts, the inhomogeneous case resembles the effects of increased nuclear spin polarization. There is almost no difference between the linear and random inhomogeneous shifts but the effect is not observed for homogeneous quadrupolar shift.}
\end{figure}
Fig. \ref{hom_vs_inh} shows the effect of homogenous and inhomogeneous quadrupolar shifts on the electron coherence time and asymptotic fidelity, which in both cases increases with increasing QP strengths $\nu_Q$. Furthermore, we note that the inhomogeneous quadrupolar shifts lead to marginally longer electron coherence times than the homogeneous ones.
%Another feature is the smaller change in coherence time in the inhomogeneous case at low quadrupolar strengths. We attribute this to the fact that there are parts of the nuclear system that have very small quadrupolar shifts, given by the terms of small $n$ in Eq. \ref{inhham}.
%\begin{figure}[h]
%\includegraphics[scale=1]{asympt}
%\caption{\label{asympt}The asymptotic fidelity for homogeneous (red curve) and inhomogeneous (green curve) quadrupolar shifts. At higher quadrupolar shifts, the inhomogeneous case resembles the effects of increased nuclear spin polarization. This effect is not observed for homogeneous quadrupolar shift.}
%\end{figure}
Finally we note that there also is a difference in the asymptotic value $F_\infty$ between the homogeneous and inhomogeneous case. For the inhomogeneous QP shifts the asymptotic value increases, an effect that is almost absent for the homogeneous case. There is, however,  not a large difference between the linear and random inhomogeneous QP shifts. Comparing to the effect of inhomogeneous quadrupolar moments to the one of increased nuclear polarization without QP shifts shown in Fig. \ref{pols1}, we find similar coherence times and fidelity asymptote at $\eta = 70$\% and $\nu_Q = 2$ kHz, suggesting both can be used as a way of increasing electron coherence time. This supports the idea that quadrupole moments may be used to obtain a quantum dot with a frozen nuclear bath, as proposed recently\cite{chekh1}.
\subsection{Combined effect of polarization and quadrupolar shift}
\begin{figure}[t]
\includegraphics[scale=1]{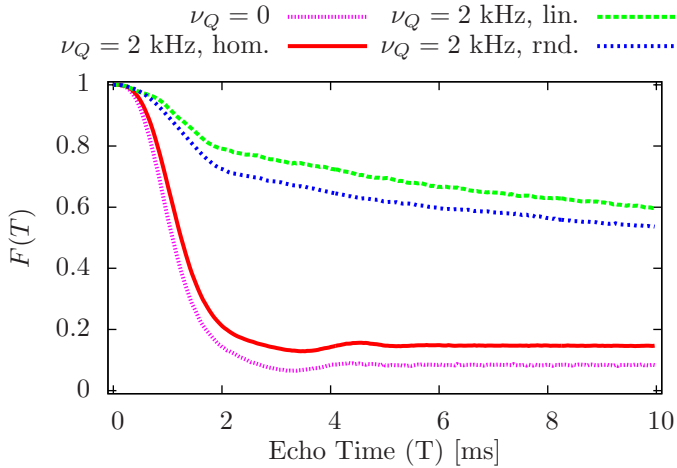}
\caption{\label{QPandPol}Echo fidelity for a nuclear spin ensemble of polarization $\eta=0.7$ with linear (green curve) and random (blue) inhomogeneous quadrupolar shifts as well as with homogenous (red curve) QP shift and without quadrupolar shift (purple curve). The inclusion of inhomogeneous qudrupolar shifts leads to a significant change in the coherence of the electron. There is a clear change in slope around $T = 2$ ms corresponding to the transitions between the two different time scales. The homogeneous quadrupolar shifts do not have a large effect on the electron coherence.}
\end{figure}

\begin{figure}
\includegraphics[scale=1]{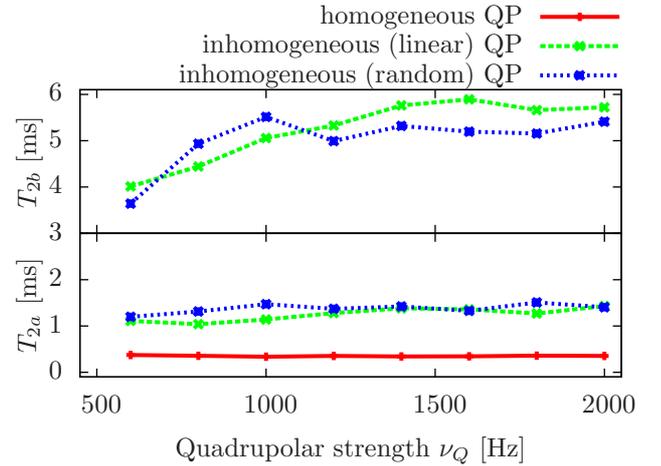}
\caption{\label{QPandPolParams}The two coherence times $T_{2a}$ and $T_{2b}$ for homogeneous (red curve), linear inhomogeneous (green curve) and random inhomogeneous (blue curve) quadrupolar shifts including $\eta = 70\%$ nuclear polarization. \textit{Upper panel:} The slow decoherence $T_{2b}$ corresponding to inhibited transitions which strongly increases with increasing quadrupolar strength $\nu_Q$. There is little difference between linear and random inhomogeneous distribution. For the homogenous QP distribution, there is no observed slow decoherence, and no $T_{2b}$ can be found. \textit{Lower panel:} The fast decoherence $T_{2a}$ corresponding to the $-1/2 \leftrightarrow 1/2$ which remains relatively constant when the quadrupolar strength is increased.}
\end{figure}

\begin{figure}
\includegraphics[scale=1]{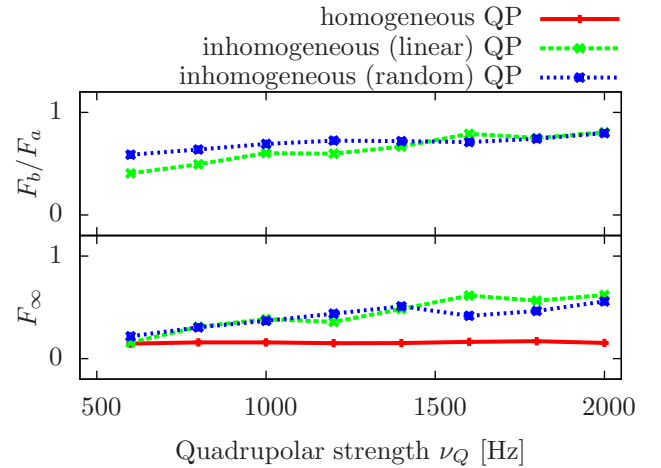}
\caption{\label{QPandPolParamsF}The coherence weights $F_a$, $F_b$, and $F_\infty$ for difference quadrupolar strengths and distributions when using $\eta = 70\%$ nuclear polarization. \textit{The upper panel} shows the ratio $F_b/F_a$ between slow and fast electron decoherence. For increasing quadrupolar strength, there is a small increase in the slow part, but little difference between linear and random quadrupolar shift distribution. \textit{The lower panel} shows the asymptotic fidelity $F_\infty$. For both distributions of inhomogeneous quadrupolar shifts, there is a clear increase with increasing quadrupolar polar strength while the homogeneous distribution remains practically constant.}
\end{figure}

When both inhomogeneous quadrupolar shifts and nuclear spin polarization are included, we expect to see
%an even larger than simply the sum of including each part separately.
further enhancement of the electron spin echo.
For a partial polarization, the population of nuclear spins will be dominated by $I_z = 3/2$ and $I_z = 1/2$ states. On the other hand, inhomogeneous quadrupolar shifts effectively suppress transitions between these states and the nuclear spins should remain mostly static. Fig. \ref{QPandPol} shows the echo fidelity $F(T)$ when quadrupolar shifts are introduced to an ensemble of nuclear spins with 70\% polarization. For homogeneous QP shifts there is little change in the electron coherence but for inhomogeneous QP shifts, the electron spin coherence is strongly increased to levels above the one corresponding to $\nu_Q = 0$ and a degree of polarization of $\eta = 90$\%, shown in Fig \ref{hom_vs_inh}. We also observe that there seems to be two time scales for the decay of the electron spin coherence. For this reason we extend the fitting function to
\begin{equation}
f(T) = F_ae^{-T^4/T_{2a}^4} + F_be^{-T^4/T_{2b}^4} + F_{\infty},
\end{equation}
where $F_a + F_b + F_{\infty} = 1$. Here, $T_{2a}$ corresponds to the decoherence of the part of the nuclear ensemble fluctuating rapidly by the unsuppressed $-1/2 \leftrightarrow 1/2$ transitions and $T_{2b}$ corresponds to the slowly fluctuating part exchanging spin via the inhibited transitions.
%
%The role of $T_{2a}$ and $T_{2b}$ is not unique, but can be interchanged.
%
The two coherence times $T_{2a}$ and $T_{2b}$ are shown as functions of the quadrupolar shift in Fig. \ref{QPandPolParams}. $T_{2b}$ increases strongly for increasing quadrupolar shift, supporting the claim that this is related to the population undergoing inhibited transitions, while $T_{2a}$ is largely unaffected by the QP shifts which indicates that this is caused by the allowed spin $I_z = 1/2 \leftrightarrow I_z = -1/2$ transitions. The effect on the asymptotic fidelity $F_\infty$ can be seen in Fig. \ref{QPandPolParamsF}. For both linear and random quadrupolar distribution $F_\infty$ increases as a function of $\nu_Q$ and reaches values over 50\% similar to the ones found for a polarization of $\eta = 90\%$ for the case of $\nu_Q = 0$, shown in Fig. \ref{pols1}. The ratio $F_b/F_a$ shows a weak increase with $\nu_Q$, indicating that the slow decoherence increase in relative magnitude to the fast one. Together, the increasing $F_\infty$ and $F_b/F_a$ demonstrate a simultaneous reduction of decoherence rates and an increase in final coherence.  For weak quadrupolar shifts $\nu_Q < 600$ Hz the reduction of nuclear spin transfer is too small for $F_b$ and $T_{2b}$ to be accurately distinguished and determined.
\section{Discussion and Conclusions}
We have investigated the effect of nuclear quadrupole moments (QPMs) on the coherence time $T_2$ of an electron in a quantum dot undergoing an electron spin echo. We found that the presence of QPMs together with electric field gradients increase the electron coherence time. The effect is larger if inhomogeneous quadrupolar shifts are present than in the case of homogeneous shifts. For the inhomogeneous case, the effect on the electron spin coherence is similar to that of increased nuclear spin polarization, suggested as an alternative method to prolong electron coherence. We found almost no difference between the two investigated distributions of quadrupolar shifts (linear and random). The impact of the QPMs is significantly increased if the nuclear spin ensemble is also partially polarized, leading to a greater population of the nuclear spin states which can only transfer spin via inhibited processes. This suggests applying the existing technique of partially polarizing the nuclear spins dynamically to quantum dots having a large built-in or externally applied inhomogeneous strain, which would lead to a significant increase of electron coherence times not achievable using only dynamic nuclear spin polarization with existing methods. Our findings also support recent suggestions\cite{chekh1} to utilize the QPMs to create a quantum dot nearly free from nuclear spin fluctuations.

\section*{Acknowledgments}
We acknowledge funding from the Konstanz Center of Applied Photonics (CAP), BMBF under the program QuaHL-Rep and from the European Union through Marie Curie ITN S$^3$NANO. E.A.C. was supported by a University of Sheffield Vice-Chancellor's Fellowship.


\begin{thebibliography}{99}
\bibitem{lossdivincenzo}
D. Loss and D. P. DiVincenzo, Phys. Rev. A \textbf{57}, 120 (1998).

\bibitem{erlingsson}
S. I. Erlingsson \textit{et al.}, Phys. Rev. B \textbf{64}, 195306 (2001).

\bibitem{merkulov}
I. A. Merkulov  \textit{et al.}, Phys. Rev. B \textbf{65}, 205309 (2002).

\bibitem{khaetskii}
A. V. Khaetskii \textit{et al.}, Phys. Rev. Lett. \textbf{88}, 186802 (2002).

\bibitem{coishbaugh}
W. A. Coish and J. Baugh, Phys. Stat. Sol. B \textbf{246}, No. 10, 2203-2215 (2009).

%\bibitem{fischer1}
%J. Fischer \textit{et al.}, Phys. Rev. B \textbf{78}, 155329 (2008).

\bibitem{petta1}
J. R. Petta \textit{et al.}, Science, \textbf{309}, 2180 (2005).

\bibitem{coish1}
W. A. Coish and D. Loss, Phys. Rev. B \textbf{70}, 195340 (2004).

\bibitem{fischer}
J. Fischer \textit{et al.}, Sol. Stat. Comm. \textbf{149} 1443-1450 (2009).

\bibitem{bayer2}
H. Kurtze \textit{et al.}, Phys. Rev. B \textbf{85}, 195303 (2012).

\bibitem{wang}
X. J. Wang \textit{et al.}, Phys. Rev. Lett. \textbf{109}, 237601 (2012).

\bibitem{slichter}
C. P. Slichter, \textit{Principles of Magnetic Resonance}, Springer (1990).

\bibitem{abragam}
A. Abragam, \textit{Principles of Nuclear Magnetism}, Oxford (1961).

\bibitem{yao}
W. Yao \textit{et al.}, Phys. Rev. B \textbf{74}, 195301 (2006).

\bibitem{economou}
S. E. Economou and E. Barnes, Phys. Rev. B \textbf{89}, 165301 (2014).

\bibitem{witzel1}
W. M. Witzel and S. Das Sarma, Phys. Rev. B \textbf{74}, 035322 (2006).

\bibitem{witzel2}
W. M. Witzel and S. Das Sarma, Phys. Rev. Lett. \textbf{98}, 077601 (2007).

\bibitem{stepanenko}
D. Stepanenko \textit{et al.}, Phys. Rev. Lett. \textbf{96}, 136401 (2006).

\bibitem{ribeiro}
H. Ribeiro and G. Burkard, Phys. Rev. Lett. \textbf{102}, 216802 (2009).

\bibitem{lee}
B. Lee \textit{et al.}, Phys. Rev. Lett. \textbf{100}, 160505 (2008).

\bibitem{khod}
K. Khodjasteh and D. A. Lidar, Phys. Rev. A \textbf{75}, 062310 (2007).

\bibitem{uhrig}
G. S. Uhrig, Phys. Rev. Lett. \textbf{98}, 100504 (2007).

\bibitem{taka}
S. Takahashi \textit{et al.}, Phys. Rev. Lett. \textbf{101}, 047601 (2008).

\bibitem{warburton}
C. Kloeffel \textit{et al.}, Phys. Rev. Lett. \textbf{106}, 046802 (2011).

\bibitem{bluhm1}
H. Bluhm \textit{et al.}, Nat. Phys. \textbf{7}, 113 (2010).

\bibitem{bluhm2}
H. Bluhm \textit{et al.}, Phys. Rev. Lett. \textbf{105}, 216803 (2010).

\bibitem{dzhioevkorenev}
R. I. Dzhioev and V. L. Korenev, Phys. Rev. Lett. \textbf{99}, 037401 (2007).

\bibitem{laird}
E. A. Laird \textit{et al.}, Phys. Rev. Lett. \textbf{97}, 056801 (2006).

\bibitem{petta2}
J. R. Petta \textit{et al.}, Phys. Rev. Lett. \textbf{100}, 067601 (2008).

\bibitem{chekh2}
E. A. Chekhovich \textit{et al.}, Phys. Rev. Lett. \textbf{104}, 066804 (2010).

\bibitem{bayer4}
R. V. Cherbunin \textit{et al.}, Phys. Rev. B \textbf{84}, 041304(R) (2011).

\bibitem{bayer3}
S. Y. Verbin \textit{et al.}, J. Exp. Theor. Phys \textbf{114}, 681 (2012).

\bibitem{blatt}
J. M. Blatt and V. F. Weisskopf, \textit{Theoretical Nuclear Physics}, Springer (1979).

\bibitem{mandel}
Mandel, L. and Wolf, E. \textit{Optical Coherence and Quantum Optics}, Cambridge (1995).


\bibitem{varwig}
S. Varwig \textit{et al.}, Phys. Rev. B \textbf{87}, 115307 (2013).

\bibitem{chekh1}
E. A. Chekhovich \textit{et al.}, arXiv:1403.1510 (2014).

\bibitem{guerrierharley}
D. J. Guerrier and R. T. Harley, Appl. Phys. Lett. \textbf{70}, 1741 (1997).

\bibitem{maletinsky}
P. Maletinsky \textit{et al.}, Nat. Phys. \textbf{5}, 407 (2009).

\bibitem{sinitsyn}
N. A. Sinitsyn \textit{et al.}, Phys. Rev. Lett. \textbf{109}, 166605 (2012).

\bibitem{flis}
K. Flisinski \textit{et al.}, Phys. Rev. B \textbf{82}, 081308(R) (2010).

\bibitem{bulutay}
C. Bulutay, Phys. Rev. B \textbf{85}, 115313 (2012).

\bibitem{bayer1}
M. S. Kuznetsova \textit{et al.}, Phys. Rev. B \textbf{89}, 125304 (2014).

\bibitem{ithier}
G. Ithier \textit{et al.}, Phys. Rev. B \textbf{72}, 134519 (2005).

\end{thebibliography}
\end{document}